\documentclass[12pt,a4paper]{article}

\usepackage{amsmath,amssymb,amsthm,natbib,graphicx,mathrsfs,tikz,amsfonts}
\usepackage{multirow}
\usepackage[colorlinks,citecolor=blue,urlcolor=blue]{hyperref}

\usetikzlibrary[decorations.pathreplacing]
\usetikzlibrary{patterns}

\begin{document}

\title{Regularized Solutions to Linear Rational Expectations Models\footnote{Thanks are due to Thomas Lubik, Davide Debortoli and Fei Tan for helpful comments and suggestions.}}
\author{Majid M.\ Al-Sadoon\\Durham University Business School}

\newtheorem{lem}{Lemma}
\newtheorem{thm}{Theorem}
\newtheorem{cor}{Corollary}
\newtheorem{prop}{Proposition}
\newtheorem{exmp}{Example}
\newtheorem{defn}{Definition}
\newtheorem{claim}{Claim}
\newtheorem{alg}{Algorithm}
\newtheorem{rmk}{Remark}
\newtheorem{res}{Result}

\renewcommand{\labelenumi}{(\roman{enumi})}

\maketitle

\abstract{
This paper proposes an algorithm for computing regularized solutions to linear rational expectations models. The algorithm allows for regularization cross-sectionally as well as across frequencies. A variety of numerical examples illustrate the advantage of regularization.

\bigskip


\bigskip

}

\newpage

\section{Introduction}
Recently, \cite{spectral} showed that non-unique solutions to multivariate linear rational expectations models (LREMs) are not generally continuous with respect to their parameters, invalidating crucial assumptions for both frequentis and Bayesian methods. For frequentist analyses, the objective function (e.g.\ the likelihood function) has to be at least continuous. For Bayesian analysis, the posterior cannot have atoms at unknown locations. \cite{spectral} demonstrated that these two conditions are not guaranteed under current methodology and proposed a regularization solution.

Regularization is a method for selecting from among infinitely many solutions to an LREM a unique solution that accords with prior information that the researcher may have about what a solution should look like (e.g. that its spectral density should concentrate in the range of observed business cycles). \cite{spectral} provided a theoretical analysis of regularizaiton. The aim of this paper is to provide an algorithm for computing such solutions based on the \cite{sims} framework.

This work is related to several more recent works. The main result of this paper builds on \cite{sunspots} and \cite{linsys}. \cite{farmeretal} and \cite{bianchinicolo} provide alternative parametrizations of solutions to LREMs to \cite{sunspots}. \cite{funovits} counts the dimension of the solution space to a given LREM. This paper can also be seen as part of the recent interest in frequency domain analysis of LREMs as exemplified by \cite{onatski}, \cite{tanwalker}, and \cite{tan}. Such methods have found important applications in addressing the identification problem for LREMs as seen in \cite{kn}, \cite{tq17}, \cite{kk18}, and \cite{ident}.

This paper is organized as follows. Section \ref{sec:review} reviews results of \cite{sims} and \cite{sunspots}. Section \ref{sec:regularization} shows how regularization can be achieved and provides the main result of this paper. Section \ref{sec:examples} provides illustrative examples of how regularization works. Section \ref{sec:conclusion} concludes. The Matlab code for reproducing the computations presented in this paper can be found in the accompanying file, \texttt{regular.zip}.

\section{Review}\label{sec:review}
We begin by reviewing results developed by \cite{sims} and \cite{sunspots}. This is necessary in order to set the notation and obtain the basic ingredients that we will need. Because regularization is only defined in a stationary context, we will restrict attention to covariance stationary solutions.

\begin{defn}
Given $(\Gamma_0,\Gamma_1,\Psi,\Pi)\in\mathbb{R}^{n\times n}\times \mathbb{R}^{n\times n}\times \mathbb{R}^{n\times l}\times \mathbb{R}^{n\times k}$, an $l$-dimensional i.i.d.\ process $z$ of mean zero and finite and positive definite variance matrix, and the formal LREM
\begin{align}
\Gamma_0 y(t)=\Gamma_1y(t-1)+\Psi z(t)+\Pi\eta(t),\qquad t\in\mathbb{Z},\label{eq:lrem}
\end{align}
a solution to \eqref{eq:lrem} is a pair $(y,\eta)$ such that:
\begin{enumerate}
\item $y$ is an $n$-dimensional process such that $y(t)$ is measurable with respect to $z(t),z(t-1),\ldots$ for all $t\in\mathbb{Z}$.
\item $\eta$ is a $k$-dimensional martingale difference sequence with respect to $z$. That is, $\eta(t)$ is measurable with respect to $z(t),z(t-1),\ldots$ and $E_t\eta(t+1)=0$ almost surely for all $t\in\mathbb{Z}$, where $E_t(\;\cdot\;)=E(\;\cdot\;|z(t),z(t-1),\ldots)$.
\item The process $(z,y,\eta)$ is jointly covariance stationary.
\item The pair satisfies equations \eqref{eq:lrem} almost surely.
\end{enumerate}
A solution $(y,\eta)$ is unique if for every other solution $(\tilde y,\tilde\eta)$, $y(t)=\tilde y(t)$ almost surely for all $t\in\mathbb{Z}$. (For ease of exposition, we will drop the ``almost surely'' in the subsequent analysis).
\end{defn}

Assuming, as \cite{sims} does, that $\det(\Gamma_0+\Gamma_1 x)$ is not identically zero (i.e.\ it is impossible to cancel out $y$ by elementary algebraic operations), then by Theorem VI.1.9 and Exercise VI.1.3 of \cite{stewartsun}, there are orthogonal matrices $Q,Z\in\mathbb{R}^{n\times n}$ such that $Q\Gamma_0 Z$ and $Q\Gamma_1 Z$ are block upper triangular with either $1\times1$ or $2\times2$ blocks on the diagonal. Under the stronger assumption that $\det(\Gamma_0+\Gamma_1 x)\neq0$ for all $x\in\mathbb{C}$ with $|x|=1$ (i.e.\ the aforementioned cancellation is impossible and there are no unit roots in the system), then these matrices can be partitioned conformably as
\begin{align}
Q\Gamma_0 Z&=\left[\begin{matrix} \Lambda_{11} & \Lambda_{12} \\ 0 & \Lambda_{22} \end{matrix}\right],& Q\Gamma_1 Z&=\left[\begin{matrix} \Omega_{11} & \Omega_{12} \\ 0 & \Omega_{22} \end{matrix}\right], \label{eq:QZ}
\end{align}
where the polynomial $\det(\Lambda_{11}+\Omega_{11}x)$ has all its zeros outside the unit circle (this implies that $\Lambda_{11}$ is non-singular), and the polynomial $\det(\Lambda_{22}+\Omega_{22}x)$ has all its zeros inside the unit circle (this implies that $\Omega_{22}$ is non-singular). As shown in the online appendix to \cite{linsys}, this step is an implicit Wiener-Hopf factorization. Note that \cite{sims} and \cite{sunspots} use the complex QZ decomposition but never explain how the final answer is real; using the real QZ decomposition obviates any need for such a discussion.

Now suppose $(y,\eta)$ is a solution to \eqref{eq:lrem}, define $w(t)=Z'y(t)$, and rewrite the system as
\begin{align*}
\Lambda w(t)=\Omega w(t-1)+Q\Psi z(t)+Q\Pi\eta(t),\qquad t\in\mathbb{Z},
\end{align*}
If we partition
\begin{align*}
w(t)=\left[\begin{array}{c} w_1(t)\\ w_2(t)\end{array}\right]
\end{align*}
conformably with \eqref{eq:QZ}, then
\begin{align}
\Lambda_{22}w_2(t)=\Omega_{22}w_2(t-1)+Q_{2\cdot}\Psi z(t)+Q_{2\cdot}\Pi\eta(t),\qquad t\in\mathbb{Z},\label{eq:w2}
\end{align}
where
\begin{align*}
Q=\left[\begin{array}{c} Q_{1\cdot}\\ Q_{2\cdot}\end{array}\right]
\end{align*}
is partitioned conformably with \eqref{eq:QZ}. Applying the conditional expectation $E_{t-1}$ we obtain
\begin{align*}
w_2(t-1)=\Omega_{22}^{-1}\Lambda_{22}E_{t-1}w_2(t),\qquad t\in\mathbb{Z},
\end{align*}
This implies that
\begin{align*}
w_2(t)=(\Omega_{22}^{-1}\Lambda_{22})^{s-t}E_tw_2(s),\qquad s\geq t.
\end{align*}
Therefore,
\begin{align*}
E\|w_2(t)\|^2&\leq\|(\Omega_{22}^{-1}\Lambda_{22})^{s-t}\|^2\;E\|E_tw_2(s)\|^2\leq\|(\Omega_{22}^{-1}\Lambda_{22})^{s-t}\|^2\;E\|w_2(s)\|^2,\qquad s\geq t.
\end{align*}
where we have used the fact that $E\|E_tw_2(s)\|^2\leq E(E_t\|w_2(s)\|^2)=E\|w_2(s)\|^2$ \cite[Theorem 9.7]{williams}. The covariance stationarity of $y$ implies that $E\|w_2(t)\|^2=E\|w_2(s)\|^2$. Since our choice of QZ decomposition ensures that the eigenvalues of $\Omega_{22}^{-1}\Lambda_{22}$ are inside the unit circle, $\|(\Omega_{22}^{-1}\Lambda_{22})^{s-t}\|<1$ for large enough $s-t$ and then it must be the case that $E\|w_2(t)\|^2=E\|w_2(s)\|^2=0$. Therefore,
\begin{align*}
w_2(t)=0,\qquad t\in\mathbb{Z}.
\end{align*}
Now plugging this back into \eqref{eq:w2} we have that
\begin{align*}
Q_{2\cdot}\Psi z(t)+Q_{2\cdot}\Pi\eta(t)=0,\qquad t\in\mathbb{Z}.
\end{align*}
Multiplying on the right by $z'(t)$, taking expectations, and utilizing the joint covariance stationarity of $\eta$ and $z$, we arrive at
\begin{align*}
Q_{2\cdot}\Psi E(z(0)z'(0))+Q_{2\cdot}\Pi E(\eta(0)z'(0))=0.
\end{align*}
But since $E(z(0)z'(0))$ is invertible by assumption, a necessary condition for existence is
\begin{align}
\mathrm{im}(Q_{2\cdot}\Psi)\subseteq\mathrm{im}(Q_{2\cdot}\Pi).\label{eq:existence}
\end{align}
It also follows that
\begin{align*}
(Q_{2\cdot}\Pi)^\dag Q_{2\cdot}\Psi z(t)+\eta(t)\in\ker(Q_{2\cdot}\Pi),\qquad t\in\mathbb{Z},
\end{align*}
where $(Q_{2\cdot}\Pi)^\dag$ is the Moore-Penrose generalized inverse of $Q_{2\cdot}\Pi$, and
\begin{align*}
E_{t-1}\left((Q_{2\cdot}\Pi)^\dag Q_{2\cdot}\Psi z(t)+\eta(t)\right)=0,\qquad t\in\mathbb{Z}.
\end{align*}
Thus, for a given matrix $K$ whose columns form a basis for $\ker(Q_{2\cdot}\Pi)$ there is a martingale difference sequence with respect to $z$, denoted by $\nu$, such that
\begin{align*}
K\nu(t)=(Q_{2\cdot}\Pi)^\dag Q_{2\cdot}\Psi z(t)+\eta(t), \qquad t\in\mathbb{Z}.
\end{align*}

Every solution is therefore representable as
\begin{align}
y(t)=\Theta_1y(t-1)+\Theta_zz(t)+\Theta_\nu\nu(t),& & 
\eta(t)=K\nu(t)-(Q_{2\cdot}\Pi)^\dag Q_{2\cdot}\Psi z(t),& & t\in\mathbb{Z},\label{eq:general}
\end{align}
with
\begin{align*}
\Theta_1=Z\left[\begin{smallmatrix}
\Lambda_{11}^{-1}\Omega_{11} & 0 \\
0 & 0
\end{smallmatrix}\right]Z',& &
\Theta_z=Z\left[\begin{smallmatrix}
\Lambda_{11}^{-1}\left(Q_{1\cdot}\Psi-Q_{1\cdot}\Pi (Q_{2\cdot}\Pi)^\dag Q_{2\cdot}\Psi \right)\\
0
\end{smallmatrix}\right],& &
\Theta_\nu=Z\left[\begin{smallmatrix}
\Lambda_{11}^{-1}Q_{1\cdot}\Pi \\
0
\end{smallmatrix}\right]K.
\end{align*}
Note that $\nu$ inters into the system along $\mathrm{rank}(Q_{1\cdot}\Pi K)$ independent directions, what \cite{funovits} calls the dimension of indeterminacy.

In fact, \eqref{eq:existence} is not only necessary but also sufficient for existence. To see this, simply construct the pair $(y,\eta)$ from \eqref{eq:general} with $\nu$ set to the zero process; it is easily checked that this pair is a solution to \eqref{eq:lrem}.

Turning now to uniqueness, we see that the arbitrary $\nu$ plays no role in the solution if and only if $\Theta_\nu=0$ or, equivalently, if and only if $Q_{\cdot1}\Pi K=0$, which can be expressed as
\begin{align}
\ker(Q_{2\cdot}\Pi)\subseteq\ker(Q_{1\cdot}\Pi).\label{eq:uniqueness}
\end{align}
Since \eqref{eq:general}, generated with $\nu(t)=Az(t)$ for $t\in\mathbb{Z}$ defines a solution for any matrix $A$, it must be that \eqref{eq:uniqueness} is necessary and sufficient for uniqueness. Note that \cite{sims} expresses \eqref{eq:uniqueness} equivalently in terms of the row spaces of $Q_{1\cdot}\Pi$ and $Q_{2\cdot}\Pi$.

To summarize, we have proven the following.

\begin{thm}[\cite{sims}]
Let $\det(\Gamma_0+\Gamma_1 x)\neq0$ for all $x\in\mathbb{C}$ with $|x|=1$. A solution to \eqref{eq:lrem} exists if and only if \eqref{eq:existence} holds. A solution is unique if and only if \eqref{eq:uniqueness} holds.
\end{thm}

\section{Regularization}\label{sec:regularization}
Current methodology utilizes \eqref{eq:general} or variants thereof. \cite{spectral} has demonstrated that these solutions can be discontinuous (as we will see shortly) and proposed using regularized solutions to ensure continuity. We now turn to the problem of computing such solutions.

We begin with the basic setting. Suppose a symmetric positive semi-definite matrix $W\in\mathbb{R}^{n\times n}$ is given and we are interested in selecting among all solutions to \eqref{eq:lrem}, one that minimizes
\begin{align*}
E\|W^{1/2}y(0)\|^2=E(y'(0)Wy(0))=\mathrm{tr}\left(WE(y(0)y'(0))\right).
\end{align*}
If the solution to \eqref{eq:lrem} is unique, there is nothing to solve for. If not, it will be convenient in the subsequent computations to introduce the martingale difference sequence, $\zeta$, defined as
\begin{align*}
\nu(t)=Bz(t)+\zeta(t),\qquad E(z(t)\zeta(t))=0,\qquad t\in\mathbb{Z}.
\end{align*}
The process $\zeta$ is the residual from regressing $\nu(t)$ on $z(t)$. This implies that
\begin{align*}
E(y(0)y'(0))=\sum_{j=0}^\infty \Theta_1^j\left(\Theta_z\Sigma_{zz}\Theta_z'+\Theta_z\Sigma_{zz}B'\Theta_\nu'+\Theta_\nu B\Sigma_{zz}\Theta_z'+\Theta_\nu B\Sigma_{zz}B'\Theta_\nu'+\Theta_\nu CC'\Theta_\nu'\right)\Theta_1^j{}',
\end{align*}
where $CC'=E(\zeta(0)\zeta'(0))$ and $\Sigma_{zz}=E(z(0)z'(0))$. Thus, finding a regularized solution is equivalent to minimizing
\begin{align*}
\mathscr{L}=\frac{1}{2}\mathrm{tr}\left(W\sum_{j=0}^\infty \Theta_1^j\left(\Theta_z\Sigma_{zz}\Theta_z'+\Theta_z\Sigma_{zz}B'\Theta_\nu'+\Theta_\nu B\Sigma_{zz}\Theta_z'+\Theta_\nu B\Sigma_{zz}B'\Theta_\nu'+\Theta_\nu CC'\Theta_\nu'\right)\Theta_1^j{}'\right)
\end{align*}
with respect to $B$ and $C$. Using the properties of the trace of a produce of matrices,
\begin{align*}
\mathscr{L}=\frac{1}{2}\mathrm{tr}\left(\left(\Theta_z\Sigma_{zz}\Theta_z'+\Theta_z\Sigma_{zz}B'\Theta_\nu'+\Theta_\nu B\Sigma_{zz}\Theta_z'+\Theta_\nu B\Sigma_{zz}B'\Theta_\nu'+\Theta_\nu CC'\Theta_\nu'\right)\Xi\right),
\end{align*}
where
\begin{align*}
\Xi=\sum_{j=0}^\infty \Theta_1^j{}'W\Theta_1^j.
\end{align*}
Note that $\Xi$ is the unique solution to the Lyapunov equation
\begin{align*}
\Xi=\Theta_1'\Xi\Theta_1+W.
\end{align*}
See Section B.1.8 of \cite{lp}. Taking the gradient of $\mathscr{L}$, we obtain the following first order conditions
\begin{align*}
\begin{aligned}
\Theta_\nu'\Xi\left(\Theta_z+\Theta_\nu B^\ast\right)&=0\\
\Theta_\nu'\Xi\Theta_\nu C^\ast&=0,
\end{aligned}
\end{align*}
If $\Theta_\nu'\Xi\Theta_\nu$ is invertible, there exists a unique regularized solution determined by
\begin{align*}
B^\ast=-(\Theta_\nu'\Xi\Theta_\nu)^{-1}\Theta_\nu'\Xi\Theta_z,\qquad C^\ast=0.
\end{align*}
If $\Theta_\nu'\Xi\Theta_\nu$ is not invertible, there are infinitely many regularized solutions determined by
\begin{align*}
B^\ast=-(\Theta_\nu'\Xi\Theta_\nu)^\dag\Theta_\nu'\Xi\Theta_z+X,\qquad C^\ast=Y,
\end{align*}
for arbitrary $X$ and $Y$ of the appropriate sizes such that $\mathrm{im}(X),\mathrm{im}(Y)\subseteq\ker(\Theta_\nu'\Xi\Theta_\nu)$.

We have established the following. First, a regularized solution to \eqref{eq:lrem} exists if and only if solutions to \eqref{eq:lrem} exist. Second, the regularized solution is unique if and only if either the solution to \eqref{eq:lrem} is unique, in which case the regularized solution is the unique solution,
\begin{align}
y(t)=\Theta_1y(t-1)+\Theta_z z(t),& & 
\eta(t)=-(Q_{2\cdot}\Pi)^\dag Q_{2\cdot}\Psi z(t),& &t\in\mathbb{Z},\label{eq:unique}
\end{align}
or $\Theta_\nu'\Xi\Theta_\nu$ is invertible, in which case the regularized solution has the representation,
\begin{align}
y(t)=\Theta_1y(t-1)+\Theta_\mathrm{reg}z(t),& & \eta(t)=
-\left(K(\Theta_\nu'\Xi\Theta_\nu)^{-1}\Theta_\nu'\Xi\Theta_z +(Q_{2\cdot}\Pi)^\dag Q_{2\cdot}\Psi \right)z(t),& &t\in\mathbb{Z},\label{eq:regular}
\end{align}
where
\begin{align*}
\Theta_\mathrm{reg}=(I-\Theta_\nu(\Theta_\nu'\Xi\Theta_\nu)^{-1}\Theta_\nu'\Xi)\Theta_z.
\end{align*}
The intuition of this result is quite simple. Write
\begin{align*}
\Theta_\nu'\Xi\Theta_\nu=\left[\begin{array}{cccc}
\Theta_\nu' W^{1/2} & \Theta_\nu'\Theta_1' W^{1/2} &  \Theta_\nu'\Theta_1^2{}' W^{1/2} & \cdots
\end{array}\right]
\left[\begin{array}{c}
W^{1/2}\Theta_\nu \\ W^{1/2}\Theta_1\Theta_\nu \\  W^{1/2}\Theta_1^2\Theta_\nu \\ \vdots
\end{array}\right].
\end{align*}
Now if $W^{1/2}\Theta_\nu$ is of full column rank, then the regularized solution is unique. That is, if $W$ attaches non-trivial weight to every contemporaneous instance of indeterminacy, then regularization eliminates indeterminacy. More generally, we have proven that regularization eliminates indeterminacy if and only if $W$ attaches non-trivial weight to every instance of indeterminacy whether contemporaneous or lagged. From a linear systems point of view, regularization leads to uniqueness if and only if the triple $(\Theta_1,\Theta_\nu,W^{1/2})$ is input observable \citep{sm}, which is to say, again, that the weight matrix detects all of the indeterminacy in the system.

The analysis above suggests a generalization of the basic setting. We have constructed an algorithm for minimizing
\begin{align*}
E\|W^{1/2}y(0)\|^2=\mathrm{tr}\left(\frac{1}{2\pi}\int Wf(\omega)d\omega\right),
\end{align*}
where $f$ is the spectral density of $y$. The above expression allows us to choose different weights along the cross-section of $y$. More generally, we may consider choosing weights on frequencies of oscillation of $y$. In particular, we may consider minimizing
\begin{align}
\mathscr{L}=\frac{1}{2}\mathrm{tr}\left(\frac{1}{2\pi}\int W(\omega)f(\omega)d\omega\right),\label{eq:L}
\end{align}
where $W$ is a bounded measurable function, with $W(\omega)$ Hermitian positive semi definite and $W(\omega)^\ast=W(-\omega)'$ for all $\omega\in(-\pi,\pi]$. If, for example, we like to impose that the solution should display the frequency characteristics of the business cycle, we could choose
\begin{align*}
W(\omega)=\begin{cases} 0,& 2\pi/32\leq|\omega|\leq 2\pi/4,\\
I,& \text{otherwise},
\end{cases}
\end{align*}
which penalizes oscillations of period smaller than a year and greater than eight years in quarterly data. To that end, we first note that
\begin{multline*}
f(\omega)=(I-\Theta_1e^{-\mathrm{i}\omega})^{-1}\big(\Theta_z\Sigma_{zz}\Theta_z'+\Theta_z\Sigma_{zz}B'\Theta_\nu'+\Theta_\nu B\Sigma_{zz}\Theta_z'\\
+\Theta_\nu B\Sigma_{zz}B'\Theta_\nu'+\Theta_\nu CC'\Theta_\nu'\big)(I-\Theta_1'e^{\mathrm{i}\omega})^{-1}.
\end{multline*}
This implies that
\begin{align*}
\mathscr{L}=\frac{1}{2}\mathrm{tr}\left(\left(\Theta_z\Sigma_{zz}\Theta_z'+\Theta_z\Sigma_{zz}B'\Theta_\nu'+\Theta_\nu B\Sigma_{zz}\Theta_z'+\Theta_\nu B\Sigma_{zz}B'\Theta_\nu'+\Theta_\nu CC'\Theta_\nu'\right)\Xi\right),
\end{align*}
where
\begin{align*}
\Xi=\frac{1}{2\pi}\int(I-\Theta_1'e^{\mathrm{i}\omega})^{-1}W(\omega)(I-\Theta_1e^{-\mathrm{i}\omega})^{-1}d\omega.
\end{align*}
It is easily checked that $\Xi$ is a real symmetric positive semi definite matrix and that it reduces to our previous expression when $W(\omega)$ is constant. Following the same line of argument as above, we arrive finally at the main result of the paper.

\begin{thm}\label{thm:regularize2}
Let $\det(\Gamma_0+\Gamma_1 x)\neq0$ for all $x\in\mathbb{C}$ with $|x|=1$. A regularized solution to \eqref{eq:lrem} that minimizes \eqref{eq:L} exists if and only if \eqref{eq:existence} holds. A regularized solution is unique if and only if either \eqref{eq:uniqueness} holds, in which case it is represented as \eqref{eq:unique}, or $\Theta_\nu'\Xi\Theta_\nu$ is invertible, in which case it is represented as \eqref{eq:regular}.
\end{thm}

\section{Examples}\label{sec:examples}
\subsection{The Cagan Model}
\begin{figure}[!ht]
\caption{Regularized Solutions to the Cagan Model.}
\label{fig:cagan}
\smallskip
\center
\includegraphics[width=12cm]{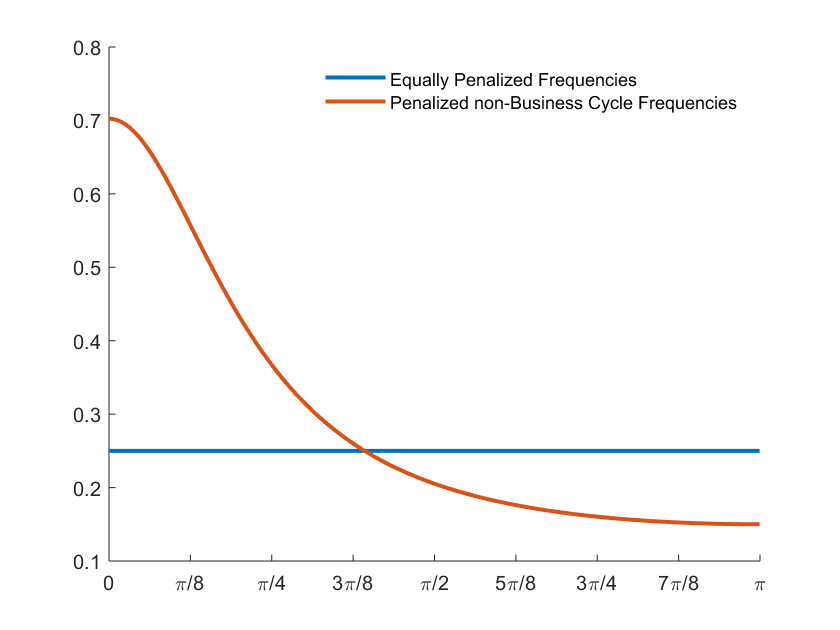}
\end{figure}

Consider first, the Cagan model with mean zero, independent, and identically distributed shocks
\begin{align*}
X_t=2E_tX_{t+1}+\varepsilon_t,\qquad t\in\mathbb{Z}.
\end{align*}
There are infinitely many solutions to this system. To compute the regularized solution minimizing $EX_t^2$, we reformulate this model as
\begin{align*}
y(t)=\left[\begin{array}{c}
X_t \\ E_tX_{t+1}
\end{array}\right],\quad
z(t)=\varepsilon_t,\quad \eta(t)=X_t-E_{t-1}X_t,\quad t\in\mathbb{Z}
\end{align*}
with
\begin{align*}
\Gamma_0=\left[\begin{array}{cc}
1 & -2\\ 1 & 0
\end{array}\right],\quad \Gamma_1=\left[\begin{array}{cc}
0 & 0\\ 0 & 1
\end{array}\right],\quad
\Psi=\left[\begin{array}{c}
1 \\ 0
\end{array}\right],\quad
\Pi=\left[\begin{array}{c}
0 \\ 1
\end{array}\right],\quad
W=\left[\begin{array}{cc}
1 & 0 \\ 0 & 0
\end{array}\right].
\end{align*}
This implies that
\begin{align*}
\Theta_1=\left[\begin{array}{cc}
         0 &   1.000\\
         0 &   0.500
\end{array}\right],\qquad \Theta_\mathrm{reg}=\left[\begin{array}{c}
    0.250\\
   -0.375
\end{array}\right].
\end{align*}
Solving for the first element, we obtain $0.5\left(\frac{0.5-L}{1-0.5L}\right)\varepsilon_t$, which was obtained analytically in \cite{spectral}. This regularized solution is actually a white noise process and therefore has a flat spectral density. We may instead impose that the solution avoid empirically unlikely frequencies. If we use the weight matrix 
\begin{align*}
W(\omega)=\begin{cases} \left[\begin{array}{cc} 0 & 0 \\ 0 & 0 \end{array}\right],& 2\pi/32\leq|\omega|\leq 2\pi/4,\vspace{0.5cm}\\
\left[\begin{array}{cc} 1 & 0 \\ 0 & 0 \end{array}\right],& \text{otherwise},
\end{cases}
\end{align*}
we obtain a different regularized solution with the spectral density plotted in the Figure \ref{fig:cagan}.

\subsection{A New Keynesian Model}
\begin{figure}
\caption{Regularized Solutions to the New Keynesian Model.}
\label{fig:nk}
\center
\includegraphics[width=16cm]{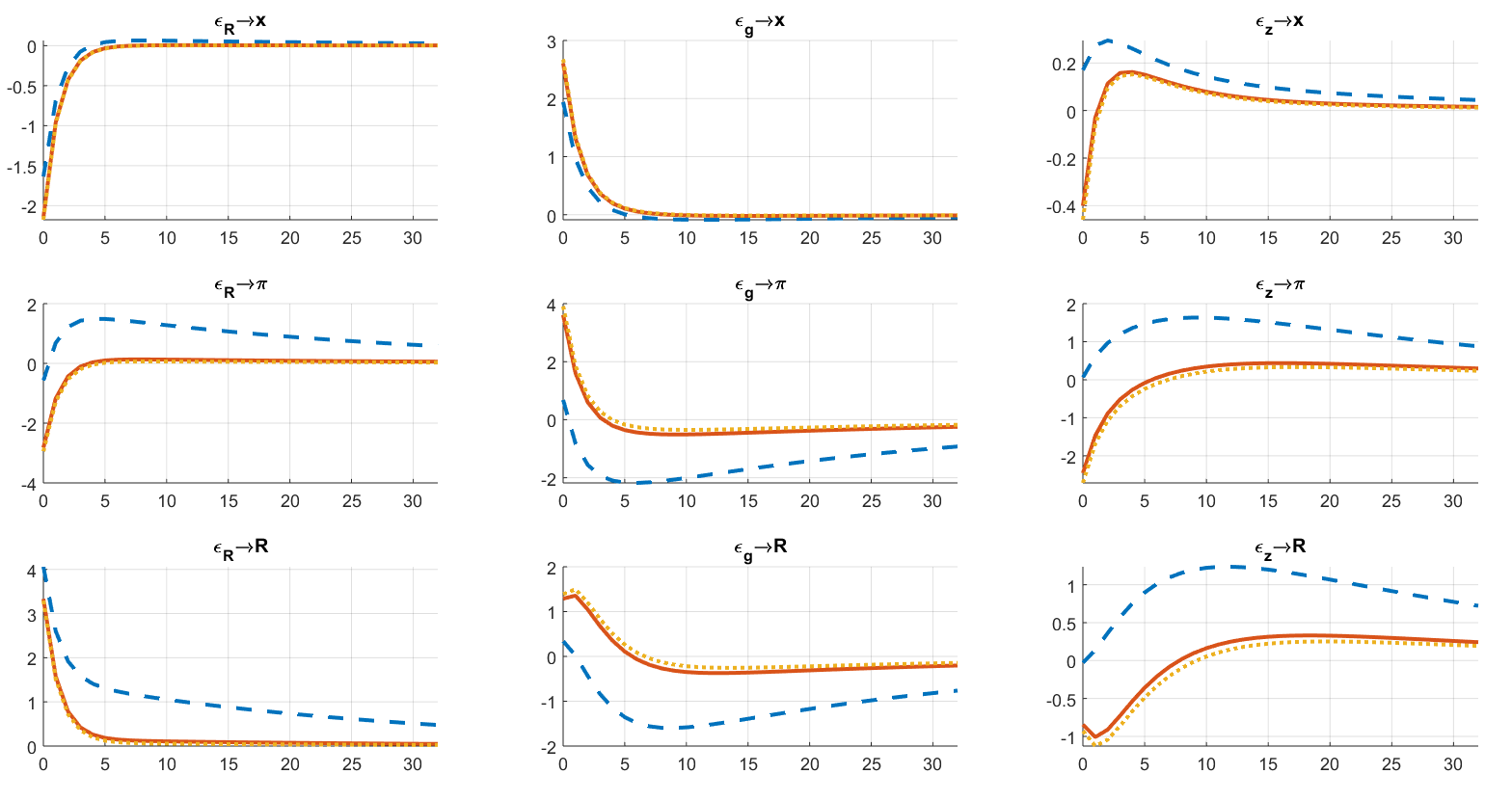}

\scriptsize{Dashed: non-regularized solution. Continuous: constant weight matrix. Dotted: variable weight matrix.}
\end{figure}

Consider next the New Keynesian model of \cite{ls04}.

\begin{gather*}
\Gamma_0=\left[\begin{array}{ccccc}
-1 & -\tau & 0 & 0 & 0 \\
0 & -\beta & 0 & 0 & 0 \\
-(1-\rho_R)\psi_2 & -(1-\rho_R)\psi_1 & 1 & 0 & (1-\rho_R)\psi_2\\
0 & 0 & 0 & 1 & 0\\
0 & 0 & 0 & 0 & 1
\end{array}\right],\quad
\Gamma_1=\left[\begin{array}{ccccc}
-1 & 0 & -\tau & 1 & 0\\
\kappa & -1 & 0 & 0 & -\kappa\\
0 & 0 & \rho_R & 0 & 0\\
0 & 0 & 0 & \rho_g & 0 \\
0 & 0 & 0 & 0 & \rho_z
\end{array}\right],\\
\Psi=\left[\begin{array}{ccc}
0 & 0 & 0 \\
0 & 0 & 0 \\
1 & 0 & 0 \\
0 & 1 & 0 \\
0 & 0 & 1
\end{array}\right],\quad
\Pi=\left[\begin{array}{cc}
-1 & 0 \\
0 & -\beta \\
0 & 0 \\
0 & 0 \\
0 & 0 
\end{array}\right].
\end{gather*}
The model is calibrated using Lubik \& Schorfheide's estimates reported in their Table 3 in the column titled ``Pre-Volcker (Prior 1)''. Figure \ref{fig:nk} plots the impulse responses of the first three variables to the three shocks. The impulse responses are generated from the non-regularized solution, the solution regularized with constant weight matrix with equal weights on the first three variables, and the solution regularized with a variable weight matrix emphasizing business cycle frequencies in the first three variables.

Clearly, regularization produces more stable dynamics. Unlike the case in Figure \ref{fig:cagan}, however, regularizing by constant and variable weight matrices did not produce dramatically different results.

\subsection{A Non-generic System}
Consider now the system
\begin{align*}
\begin{aligned}
E_tX_{1t+2}&=\varepsilon_{1t},\\
\theta E_tX_{1t+1}+X_{2t}&=\varepsilon_{2t},
\end{aligned}\qquad t\in\mathbb{Z}.
\end{align*}
The shocks are again zero mean, independent, and identically distributed. This system also has infinitely many solutions. Although it is simple, it concretely illustrates the failure of current methodology to account for discontinuity of solutions to LREMs. \cite{spectral} demonstrates its discontinuity analytically and studies its Gaussian likelihood function. We will now demonstrate its discontinuity numerically.

In order to reformulate this system into the form \eqref{eq:lrem}, we use the second equation to obtain
\begin{align*}
\theta E_tX_{1t+2}+E_tX_{2t+1}=0,\qquad t\in\mathbb{Z},
\end{align*}
and then combine this equation with the first equation of the original system to obtain
\begin{align*}
\begin{aligned}
E_tX_{2t+1}&=-\theta\varepsilon_{1t},\\
\theta E_tX_{1t+1}+X_{2t}&=\varepsilon_{2t},
\end{aligned}\qquad t\in\mathbb{Z}.
\end{align*}
This system is equivalent to the original one, provided $\theta\neq0$. We can now set
\begin{align*}
y(t)=\left[\begin{array}{c}
X_{1t} \\ X_{2t} \\ E_tX_{1t+1} \\ E_tX_{2t+1}
\end{array}\right],\quad
z(t)=\left[\begin{array}{cc}
\varepsilon_{1t}\\ \varepsilon_{2t}
\end{array}\right],\quad \eta(t)=\left[\begin{array}{cc}
X_{1t}-E_{t-1}X_{1t}\\ X_{2t}-E_{t-1}X_{2t}
\end{array}\right],\quad t\in\mathbb{Z}
\end{align*}
with
\begin{align*}
\Gamma_0=\left[\begin{array}{cccc}
0 & 0 & 0 & 1\\ 0 & 1 & \theta & 0 \\ 1 & 0 & 0 & 0 \\ 0 & 1 & 0 & 0
\end{array}\right],\quad \Gamma_1=\left[\begin{array}{cccc}
0 & 0 & 0 & 0 \\ 0 & 0 & 0 & 0 \\ 0 & 0 & 1 & 0 \\ 0 & 0 & 0 & 1
\end{array}\right],\quad
\Psi=\left[\begin{array}{cc}
-\theta & 0\\ 0 & 1\\ 0 & 0\\ 0 & 0
\end{array}\right],\quad
\Pi=\left[\begin{array}{cc}
0 & 0\\ 0 & 0 \\ 1 & 0\\ 0 & 1
\end{array}\right].
\end{align*}
The weight matrix is
\begin{align*}
W=\left[\begin{array}{cccc}
1 & 0 & 0 & 0 \\ 0 & 1 & 0 & 0 \\ 0 & 0 & 0 & 0 \\ 0 & 0 & 0 & 0
\end{array}\right].
\end{align*}

For $\theta=10^{-6}$, the first three impulse responses of the non-regularized solution are
\begin{align*}
\left[\begin{array}{cc}
0 & 0 \\ 0 & 0
\end{array}\right],\quad
\left[\begin{array}{cc}
0 & 10^6\\ -10^{-6} & 0
\end{array}\right],\quad
\left[\begin{array}{cc}
1 & 0 \\ 0 & 0
\end{array}\right].
\end{align*}
Clearly, these are quite far from the impulse responses of the $\theta=0$ model, which ought to be
\begin{align*}
\left[\begin{array}{cc} 0 & 0 \\ 0 & 1 \end{array}\right],\quad \left[\begin{array}{cc} 0 & 0 \\ 0 & 0 \end{array}\right],\quad \left[\begin{array}{cc} 1 & 0 \\ 0 & 0 \end{array}\right].
\end{align*}
On the other hand, the first three impulse responses of the regularized solution are
\begin{align*}
\left[\begin{array}{cc} 0 & -0.0002 \\ 0 & 1 \end{array}\right],\quad \left[\begin{array}{cc} 0 & 10^{-6} \\ -10^{-6} & 0 \end{array}\right],\quad \left[\begin{array}{cc} 1 & 0 \\ 0 & 0 \end{array}\right].
\end{align*}
The continuity of regularized solutions is proven in Theorem 6 of \cite{spectral}.

\section{Conclusion}\label{sec:conclusion}
This paper has provided an algorithm for computing regularized solutions to LREMs. This work suggests at least three venues for further investigation. First, it is likely that regularization helps resolve identifiability issues in LREMs due to its imposition of uniqueness but since it is strictly more general than the class of solutions considered in \cite{ident} its identifiability requires separate examination. Second, the algorithm presented here is given without any claim to efficiency; it would be helpful to consider other methods of obtaining regularized solutions and compare their accuracy and speed. Finally, recent work has sought to relax the assumption that the information set includes all exogenous variables (e.g.\ \cite{ht}, \cite{rondina}, \cite{angeletos}, and \cite{han}); regularization in that context would be a fruitful venue for follow up work.

\scriptsize
\setlength{\bibsep}{-1pt}
\bibliographystyle{newapa}
\bibliography{spectral}

\end{document}